\documentclass{PoS}
\pdfoutput=1
\usepackage{amstext}
\newcommand{\Tr}{\mathop{Tr}\nolimits}
\title{Four-loop results for the cusp anomalous dimension%
\raisebox{26mm}[0mm][0mm]{\makebox[0mm][r]{\textnormal{MITP/18-056}}}%
\raisebox{18mm}[0mm][0mm]{\makebox[0mm][r]{\textnormal{MPP-2018-167}}}}
\ShortTitle{Four-loop results for the cusp anomalous dimension}
\author{R.~Br\"{u}ser$^a$, \speaker{A.\,G.~Grozin}$^{b,c}$, J.\,M.~Henn$^{a,d}$ and M.~Stahlhofen$^a$\\
\llap{$^a$}PRISMA Cluster of Excellence,
Johannes Gutenberg-Universit\"{a}t, 55128 Mainz, Germany\\
\llap{$^b$}Budker Institute of Nuclear Physics, Novosibirsk 630090, Russia\\
\llap{$^c$}Novosibirsk State University, Novosibirsk 630090, Russia\\
\llap{$^d$}Max-Planck-Institut f\"{u}r Physik,
Werner-Heisenberg-Institut, 80805 M\"{u}nchen, Germany\\
E-mail: \email{brueser@uni-mainz.de}, \email{A.G.Grozin@inp.nsk.su}, \email{henn@uni-mainz.de} and \email{mastahlh@uni-mainz.de}}
\abstract{We review the current status of calculations
of the HQET field anomalous dimension and the cusp anomalous dimension.
In particular, we give the results at 4 loops for the quartic Casimir contribution,
and for the full QED case, up to $\varphi^6$ in the small angle expansion.
Furthermore, we discuss the leading terms in the anti-parallel lines limit at four loops.}
\FullConference{Loops and Legs in Quantum Field Theory (LL2018)\\
29 April 2018 -- 04 May 2018\\
St.~Goar, Germany}

\begin{document}

\section{Introduction}
\label{S:Intro}

We review the current status of calculations of 2 QCD quantities:
the HQET field anomalous dimension $\gamma_h$
and the cusp anomalous dimension $\Gamma_{\text{cusp}}(\varphi)$.
Due to non-abelian exponentiation,
they have only a subset of all possible color structures.
At small angles $\Gamma_{\text{cusp}}(\varphi)$ is a regular series in $\varphi^2$;
at large angles $\Gamma_{\text{cusp}}(\varphi) = K \varphi + \mathcal{O}(\varphi^0)$,
where $K$ is the light-like cusp anomalous dimension.
These quantities are known at 3 loops: \cite{Melnikov:2000zc,Chetyrkin:2003vi}
and~\cite{Grozin:2014hna,Grozin:2015kna}.
The status of 4-loop calculations is summarized in Table~\ref{T}.

\begin{table}[ht]
\begin{center}
\begin{tabular}{l|l|l|l|l}
\hline
& $\gamma_h$ & $\Gamma_{\text{cusp}}(\varphi)$ & $\varphi\ll1$ & $\varphi\gg1$ \\
\hline
$C_F (T_F n_l)^3$ & \cite{Broadhurst:1994se} & \cite{Beneke:1995pq} & & \\
\hline
$C_F^2 (T_F n_l)^2$ & \cite{Grozin:2015kna,Grozin:2016ydd} & \cite{Grozin:2015kna,Grozin:2016ydd} & & \\
$C_F C_A (T_F n_l)^2$ & \cite{Marquard:2018rwx,BGHS} & & \cite{BGHS} & \cite{Henn:2016men,Davies:2016jie} \\
\hline
$C_F^3 T_F n_l$ & \cite{Grozin:2018vdn} & \cite{Grozin:2018vdn} & & \\
$d_{FF} n_l$ & \cite{Grozin:2017css} & & \cite{Grozin:2017css} & \cite{Moch:2017uml,Moch:2018wjh}$*$ \\
$C_F^2 C_A T_F n_l$ & \cite{Marquard:2018rwx}$*$ & & & \cite{Moch:2017uml}$*$ \\
$C_F C_A^2 T_F n_l$ & \cite{Marquard:2018rwx}$*$ & & & \cite{Moch:2017uml}$*$ \\
$n_l^1$, $N_c\to\infty$ & & & & \cite{Henn:2016men,Moch:2017uml} \\
\hline
$C_F C_A^3$ & \cite{Marquard:2018rwx}$*$ & & & \cite{Moch:2017uml}$*$ \\
$d_{FA}$ & \cite{Marquard:2018rwx}$*$ & & & \cite{Moch:2017uml,Moch:2018wjh}$*$ \\
$n_l^0$, $N_c\to\infty$ & & & & \cite{Lee:2016ixa,Moch:2017uml} \\
\hline
QED & \cite{Grozin:2018vdn} & & \cite{Grozin:2018vdn} &
\end{tabular}
\end{center}
\caption{4-loop contributions to $\gamma_h$ and $\Gamma_{\text{cusp}}(\varphi)$.
The sign $*$ means that the contribution is only known numerically.}
\label{T}
\end{table}

The calculation of the $C_F C_A (T_F n_l)^2$ structure in $\Gamma_{\text{cusp}}(\varphi)$ at $\varphi\ll1$
is in progress~\cite{BGHS}.
The $C_F^3 T_F n_l$ structure is discussed in Sect.~\ref{FFFl}~\cite{Grozin:2018vdn},
and the $d_{FF} n_l$ structure (where $d_{FF} = d_F^{abcd} d_F^{abcd} / N_F$) ---
in Sect.~\ref{dFFl}~\cite{Grozin:2017css}.
Not much is known about the $C_F C_A^3$ structure;
when the Euclidean $\varphi$ is $\pi-\delta$, $\delta\to0$,
it has a $\log(\delta)/\delta$ term (Sect.~\ref{delta}~\cite{GS})
(calculation of the non-logarithmic $1/\delta$ term is much more difficult and not yet done).

\begin{figure}[ht]
\begin{center}
\includegraphics{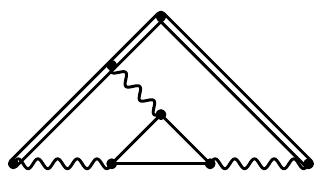}
\end{center}
\caption{A diagram for the 3-loop $C_F C_A T_F n_l$ term in $\Gamma_{\text{cusp}}(\varphi)$.}
\label{F:Conj}
\end{figure}

It has been conjectured in~\cite{Grozin:2014hna,Grozin:2015kna}
that the cusp anomalous dimension can be represented in the form
\begin{equation}
\Gamma_{\text{cusp}}(\varphi) = C_F \frac{a}{\pi} \biggl[ \Omega(\varphi) + C_A \Omega_A(\varphi) \frac{a}{\pi}
+ C_A^2 \Omega_{AA}(\varphi) \left(\frac{a}{\pi}\right)^{\!\!2} \biggr]
+ \mathcal{O}(a^4)
\label{conj}
\end{equation}
containing no $n_l$, via the effective coupling
\begin{eqnarray}
\frac{a}{\pi} &=& \frac{\alpha_s}{\pi}
+ \left(C_A B_A + T_F n_l B_l\right) \left(\frac{\alpha_s}{\pi}\right)^{\!\!2}
+ \left(C_A^2 B_{AA} + C_F T_F n_l B_{Fl} + C_A T_F n_l B_{Al} + (T_F n_l)^2 B_{ll}\right) \left(\frac{\alpha_s}{\pi}\right)^{\!\!3}
\nonumber\\
&&{} + \mathcal{O}(\alpha_s^4)
\label{conja}
\end{eqnarray}
which is determined from the condition that at $\varphi\to\infty$ the $\mathcal{O}(\varphi)$ asymptotics
is given by the first term in~(\ref{conj}).
This is true up to 3 loops.
For example, the 3-loop $C_F C_A T_F n_l$ term in $\Gamma_{\text{cusp}}(\varphi)$
(a typical diagram is shown in Fig.~\ref{F:Conj})
is a combination of 2- and 1-loop terms:
\begin{equation}
\Gamma_{\text{cusp}}(\varphi) = \cdots + C_F C_A T_F n_l
\left[B_{Al} \Omega(\varphi) + 2 B_l \Omega_A(\varphi)\right]
\left(\frac{\alpha_s}{\pi}\right)^{\!\!3} + \cdots
\label{conj3}
\end{equation}
This conjecture has been disproved for a quartic Casimir color structure~\cite{Grozin:2017css} (Sect.~\ref{dFFl}).
Remarkably, numerically the conjectured formula is very close to the exact one, cf.~Sect.~\ref{dFFl}.

\section{$C_F^3 T_F n_l$}
\label{FFFl}

\begin{figure}[ht]
\begin{center}
\begin{picture}(94,20)
\put(21,9.75){\makebox(0,0){\includegraphics{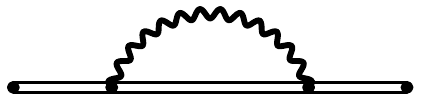}}}
\put(73,12.5){\makebox(0,0){\includegraphics{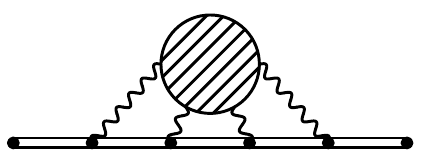}}}
\put(21,0){\makebox(0,0)[b]{a}}
\put(73,0){\makebox(0,0)[b]{b}}
\end{picture}
\end{center}
\caption{Webs: (a) 2-leg (the thick line is the full photon propagator);
(b) 4-leg (the blob is the sum of connected diagrams).}
\label{F:webs}
\end{figure}

This is a QED problem.
Due to exponentiation, the coordinate-space propagator of the Bloch--Nordsieck field
(i.\,e.\ the straight Wilson line $W$) is
\begin{equation}
W = \exp \left( \sum w_i \right)\,,
\label{gammah:exp}
\end{equation}
where $w_i$ are single-web diagrams.
Due to $C$ parity conservation in QED, webs have even numbers of legs (Fig.~\ref{F:webs}).
Webs with 4 legs (Fig.~\ref{F:webs}b) first appear at 4 loops (Sect.~\ref{dFFl}).
All contributions to $\log W$~(\ref{gammah:exp})
are gauge invariant except the 1-loop one,
because proper vertex functions with any numbers of photon legs
are gauge invariant and transverse with respect to each photon leg
due to the QED Ward identities.

In Landau gauge we obtain
\begin{equation}
\gamma_h = \frac{\alpha}{4\pi} \left[ - 6
+ n_l \sum_{L=1}^\infty \left( - 6 \bar{\Pi}_L + 2 \bar{\beta}_L\right) \left(\frac{\alpha}{4\pi}\right)^L \right]
+ (n_l^{>1} \text{ terms}) + (w_{>2 \text{ legs}} \text{ terms})\,,
\label{gammah:gen}
\end{equation}
where the photon self energy is
\begin{equation}
\Pi_L = \left(\frac{\bar{\beta}_L}{L\varepsilon} + \bar{\Pi}_L\right) n_l + (n_l^{>1} \text{ terms})\,,\quad
\beta_L = \bar{\beta}_L n_l + (n_l^{>1} \text{ terms})
\label{gammah:Pi}
\end{equation}
($\beta_L$ is the $L$-loop $\beta$ function coefficient).
Substituting $\bar{\Pi}_L$~\cite{Ruijl:2017eht},
restoring color structures and inserting the 1-loop gauge dependence,
we obtain up to 5 loops~\cite{Grozin:2018vdn}
\begin{eqnarray}
&&\gamma_h = 2 (a-3) C_F \frac{\alpha_s}{4\pi} + T_F n_l C_F \left(\frac{\alpha_s}{4\pi}\right)^2 \biggl[
\frac{32}{3}
- 6 \left(16 \zeta_3 - 17\right) C_F \frac{\alpha_s}{4\pi}
\nonumber\\
&&{} + \frac{16}{3} \left(180 \zeta_5 - 111 \zeta_3 - 35\right) \left(C_F \frac{\alpha_s}{4\pi}\right)^2
\nonumber\\
&&{} - 6 (2240 \zeta_7 - 1960 \zeta_5 - 104 \zeta_3 - 5) \left(C_F \frac{\alpha_s}{4\pi}\right)^3
+ \mathcal{O}(\alpha_s^4) \biggr]
+ (\text{other color structures})\,.
\label{gammah:gamma}
\end{eqnarray}

\begin{figure}[ht]
\begin{center}
\begin{picture}(52,26)
\put(26,12){\makebox(0,0){\includegraphics{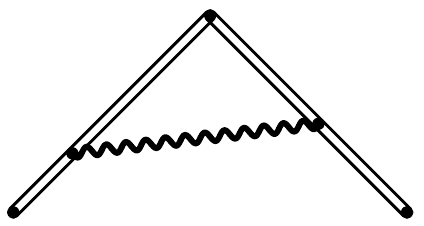}}}
\put(26,24){\makebox(0,0){$0$}}
\put(1.5,2){\makebox(0,0){$-vt$}}
\put(6.5,9){\makebox(0,0){$-vt_1$}}
\put(49,2){\makebox(0,0){$v't$}}
\put(41,12){\makebox(0,0){$v't_2$}}
\end{picture}
\end{center}
\caption{Cusp: the 2-leg webs (the thick line is the full photon propagator).}
\label{F:cusp}
\end{figure}
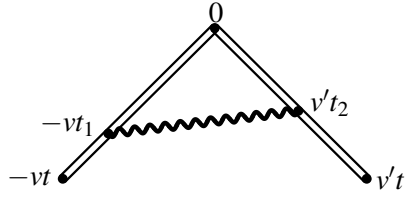

Now we consider the cusped Wilson line $W(\varphi)$
from $x=-vt$ to $0$ and then to $x'=v't$:
\begin{equation}
\log\frac{W(\varphi)}{W(0)} = \sum(w_i(\varphi)-w_i(0))\,.
\label{Gamma:exp}
\end{equation}
Diagrams in which all photon vertices are to the left (or to the right) of the cusp
cancel in $w_i(\varphi)-w_i(0)$.
The remaining 2-leg webs are represented by Fig.~\ref{F:cusp}.
At 4 loops 4-leg webs appear (Sect.~\ref{dFFl}).

The $L$-loop $n_l^1$ contribution is proportional to
\begin{equation}
\int_0^t dt_1 \int_0^t dt_2\,v_\mu v'_\nu \bar{D}_{L-1}^{\mu\nu}(vt_1+v't_2)\,,
\label{Gamma:wL}
\end{equation}
where $\bar{D}_L^{\mu\nu}$ is the $n_l^1$ term in the $L$-loop photon propagator.
Calculating the integral we obtain
\begin{equation}
\Gamma_{\text{cusp}}(\varphi) = 4 (\varphi \coth\varphi - 1) \frac{\alpha}{4\pi} \biggl[1
+ n_l \sum_{L=1}^\infty \bar{\Pi}_L \left(\frac{\alpha}{4\pi}\right)^L \biggr]
+ (n_l^{>1} \text{ terms}) + (w_{>2 \text{ legs}} \text{ terms})\,.
\label{Gamma:gen}
\end{equation}
The QCD result up to 5 loops is~\cite{Grozin:2018vdn}
\begin{eqnarray}
&&\Gamma_{\text{cusp}}(\varphi) = 4 (\varphi \coth\varphi - 1) C_F \frac{\alpha_s}{4\pi} \biggl\{1
+ T_F n_l \frac{\alpha_s}{4\pi} \biggl[
- \frac{20}{9}
+ \left(16 \zeta_3 - \frac{55}{3}\right) C_F \frac{\alpha_s}{4\pi}
\nonumber\\
&&{} - 2 \left(80 \zeta_5 - \frac{148}{3} \zeta_3 - \frac{143}{9}\right) \left(C_F \frac{\alpha_s}{4\pi}\right)^2
\nonumber\\
&&{} + \left(2240 \zeta_7 - 1960 \zeta_5 - 104 \zeta_3 + \frac{31}{3}\right) \left(C_F \frac{\alpha_s}{4\pi}\right)^3
+ \mathcal{O}(\alpha_s^4) \biggr] \biggr\}
+ (\text{other color structures})\,.
\label{Gamma:Gamma}
\end{eqnarray}

\section{$d_{FF} n_l$}
\label{dFFl}

Casimir scaling holds for $\gamma_h$ and $\Gamma_{\text{cusp}}(\varphi)$ up to 3 loops.
At 4 loops quartic Casimir color structures $d_{RF} n_l$ and $d_{RA}$ appear,
where $d_{RR'} = d_R^{abcd} d_{R'}^{abcd} / N_R$, $d_R^{abcd} = \Tr t_R^{(a} t_R^b t_R^c t_R^{d)}$, $N_R = \Tr\mathbf{1}_R$.
They cannot be represented as the quadratic Casimirs $C_R$ times a universal constant.
Therefore, Casimir scaling breaks at 4 loops,
unless by some miracle the coefficients of both quartic Casimirs identically vanish.
But they don't vanish:
in $\Gamma_{\text{cusp}}$ at Euclidean angle $\varphi_E\to\pi$ they are given by
the corresponding coefficients in the 3-loop static potential,
which are known~\cite{Anzai:2010td} and non-zero.
So, Casimir scaling breaks at 4 loops, as expected.
This breaking has been shown not to vanish in other regions of $\varphi$, too:
at Minkowsky angles $\varphi_M\gg1$
(in $\mathcal{N}=4$ SYM~\cite{Boels:2017skl,Boels:2017ftb}
and in QCD~\cite{Moch:2017uml,Moch:2018wjh})
and at $\varphi\ll1$~\cite{Grozin:2017css}.

\begin{figure}[ht]
\begin{center}
\begin{picture}(85,11.6)
\put(12.4,5.8){\makebox(0,0){\includegraphics{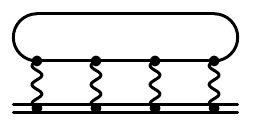}}}
\put(42.5,5.8){\makebox(0,0){\includegraphics{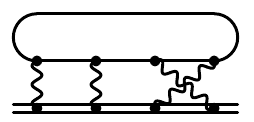}}}
\put(72.6,5.8){\makebox(0,0){\includegraphics{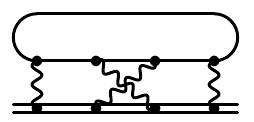}}}
\end{picture}
\end{center}
\caption{The $d_{FF} n_l$ contribution to the HQET self energy
(symmetric diagrams implied).}
\label{F:dFF}
\end{figure}

The $d_{FF} n_l$ contribution to the HQET self energy is given by 3 different diagrams (Fig.~\ref{F:dFF}).
It is gauge invariant due to QED Ward identities.
We reduce these diagrams at residual energy $\omega<0$ to master integrals,
and obtain $\varepsilon$ expansions of non-trivial master integrals
using \textsc{HyperInt}~\cite{Panzer:2014caa}.
The result is~\cite{Grozin:2017css}
\begin{equation}
\gamma_h\Big|_{d_{FF} n_l} = - d_{FF} n_l \bigg(\frac{\alpha_s}{\pi} \bigg)^{\!\!4}
\left( \frac{5}{4} \zeta_5 - \frac{2}{3} \pi^2 \zeta_3 - \zeta_3 + \frac{2}{3} \pi^2 \right)\,.
\label{HQET:gamma}
\end{equation}

\begin{figure}[ht]
\begin{center}
\begin{picture}(85,30)
\put(12.4,23.504){\makebox(0,0){\includegraphics{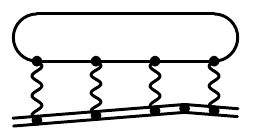}}}
\put(42.5,23.504){\makebox(0,0){\includegraphics{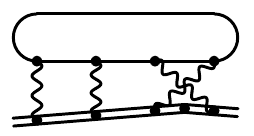}}}
\put(72.6,23.504){\makebox(0,0){\includegraphics{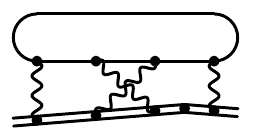}}}
\put(12.4,6.256){\makebox(0,0){\includegraphics{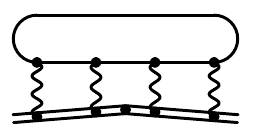}}}
\put(42.5,6.256){\makebox(0,0){\includegraphics{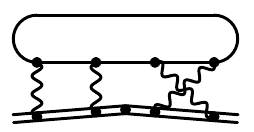}}}
\put(72.6,6.256){\makebox(0,0){\includegraphics{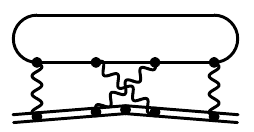}}}
\end{picture}
\end{center}
\caption{The $d_{FF} n_l$ contribution to the vertex
(symmetric diagrams implied).}
\label{F:Vertex}
\end{figure}

We have also calculated the vertex at the residual energies of its legs $\omega_1=\omega_2$ expanded in $\varphi$.
It is given by 6 different diagrams (Fig.~\ref{F:Vertex});
of course, it reduces to the same master integrals.
The result is~\cite{Grozin:2017css}
(the $\varphi^6$ term is new)
\begin{eqnarray}
&&\Gamma_{\text{cusp}}(\varphi)\Big|_{d_{FF} n_l} = d_{FF} n_l \bigg(\frac{\alpha_s}{\pi} \bigg)^{\!\!4} \frac{\varphi^2}{9} \Bigg[
\pi^2 \bigg(\!\! - 4 \zeta_3 + \frac{5}{12} \pi^2 + \frac{5}{6} \bigg)
\nonumber\\
&&{} + \varphi^2 \bigg(\!\!
- 4 \zeta_5
- \frac{16}{75} \pi^2 \zeta_3
+ \frac{71}{25} \zeta_3
+ \frac{49}{900} \pi^4
- \frac{157}{900} \pi^2
- \frac{23}{100}
\bigg)
\nonumber\\
&&{} + \varphi^4 \bigg(\!\!
- \frac{64}{147} \zeta_5
- \frac{32}{1225} \pi^2 \zeta_3
+ \frac{983}{3675} \zeta_3
+ \frac{421}{66150} \pi^4
- \frac{1333}{66150} \pi^2
+ \frac{797}{29400}
\bigg)
+ \mathcal{O}(\varphi^6) \Bigg]
\nonumber\\
&&{} = d_{FF} n_l \bigg(\frac{\alpha_s}{\pi} \bigg)^{\!\!4} \varphi^2
\big(0.150721 + 0.00965191\,\varphi^2 + 0.000925974\,\varphi^4 + \mathcal{O}(\varphi^6)\big)\,.
\label{exact}
\end{eqnarray}

The conjecture from~\cite{Grozin:2014hna,Grozin:2015kna} predicts
\begin{equation}
\Gamma_{\text{cusp}}(\varphi) = C_F \Omega(\varphi) \frac{a}{\pi} + \cdots
= \cdots + d_{FF} n_l B \Omega(\varphi) \left(\frac{\alpha_s}{\pi}\right)^{\!\!4} + \cdots\,,\quad
\frac{a}{\pi} = \cdots + \frac{d_{FF} n_l}{C_F} B \left(\frac{\alpha_s}{\pi}\right)^{\!\!4} + \cdots
\label{conjfail}
\end{equation}
The normalization factor $B$ can be found from the limit of euclidean $\varphi=\pi-\delta$, $\delta\to0$,
where the 4-loop $\Gamma_{\text{cusp}}(\varphi)$ is related to the 3-loop quark--antiquark potential which is known~\cite{Lee:2016cgz}.
This gives the prediction
\begin{eqnarray}
&&\Gamma_{\text{cusp}}(\varphi) \big|_{d_{FF} n_l}^\text{conj} = d_{FF} n_l \bigg(\frac{\alpha_s}{\pi} \bigg)^{\!\!4}
\frac{\varphi^2}{192} \biggl(1 + \frac{\varphi^2}{15} + \frac{2}{315} \varphi^4 + \mathcal{O}(\varphi^6) \biggr)
\bigg( 16 \pi ^4 \log^2 2 - 336 \pi^2 \zeta_3 \log\,2
\nonumber\\
&&{} - \frac{16}{3} \pi^4 \log\,2 - 32 \pi^2 \log\,2
+ \frac{488}{3} \pi^2 \zeta_3 - \frac{5}{3} \pi^6 + \frac{92}{3} \pi^4  - \frac{632}{9} \pi^2
\bigg)
\nonumber\\
&&{} = d_{FF} n_l \bigg(\frac{\alpha_s}{\pi} \bigg)^{\!\!4} \varphi^2
\big(0.14801 + 0.00986736\,\varphi^2 + 0.000939748\,\varphi^4 + \mathcal{O}(\varphi^6) \big)\,.
\label{conjpred}
\end{eqnarray}
So, the conjecture has been disproved.
Curiously, the numerical values~(\ref{conjpred}) of the coefficients predicted by the conjecture
are quite close to the exact ones~(\ref{exact}).

Adding~(\ref{gammah:gamma}), (\ref{HQET:gamma}) and the known contributions with higher powers of $n_l$,
we obtain the anomalous dimension of the Bloch--Nordsieck field in QED up to 4 loops,
completely analytically.
Adding~(\ref{Gamma:Gamma}), (\ref{exact}) and the known contributions with higher powers of $n_l$,
we obtain the QED cusp anomalous dimension expanded up to $\varphi^6$.

\section{$\Gamma_{\text{cusp}}(\pi-\delta)$}
\label{delta}

This Section is based on work in progress~\cite{GS}.
In Euclidean space the angle $\varphi$ varies from 0 to $\pi$.
When $\varphi=\pi-\delta$, $\delta\to0$,
the two world lines forming the cusp come together.
At 2 loops $\Gamma_{\text{cusp}}(\pi-\delta) \sim 1/\delta$,
and the coefficient is related
to the 1-loop quark--antiquark potential $V(r)$~\cite{Kilian:1993nk}.
This is explained by conformal symmetry;
in QCD it is broken by the $\beta$ function,
and at 3 loops this relation is broken by an extra term
proportional to $\beta_0$~\cite{Grozin:2014hna}.
At 4 loops a new $\log(\delta)/\delta$ term appears
(if no resummation is done).
It is similar to the 3-loop $\log(\mu r)$ term
in the static quark-antiquark potential~\cite{Brambilla:1999qa,Brambilla:1999xf}.

\begin{figure}[ht]
\begin{center}
\begin{picture}(47,20.5)
\put(23.5,12){\makebox(0,0){\includegraphics{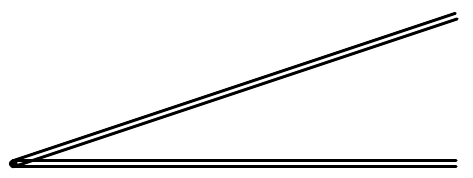}}}
\put(20,14.5){\makebox(0,0){$\vec{r} = \vec{u} t$}}
\put(20,0){\makebox(0,0)[b]{$\vec{r} = \vec{0}$}}
\end{picture}
\end{center}
\caption{The Wilson line describing production of a heavy quark--antiquark pair
with a small relative velocity $\vec{u}$.}
\label{F:L}
\end{figure}

Let's consider the cusped Wilson line in Minkowski space (Fig.~\ref{F:L}).
It is formed by the static quark and antiquark world lines
$\vec{r}=0$ and $\vec{r} = \vec{u} t$,
where $\vec{u}$ is the small relative velocity ($u=|\vec{u}|\ll1$).
At the end of calculation we'll analytically continue the result
to Euclidean space ($u = i\delta$).
We neglect all terms suppressed by powers of $u$.
It is convenient to use Coulomb gauge.
The static quark and antiquark interact by exchanging instantaneous Coulomb gluons:
\begin{equation}
V(\vec{q}) = - C_F \frac{g_0^2}{\vec{q}^2}\,,\quad
V(\vec{r}) = - C_F \kappa_0 \frac{g_0^2}{4\pi} \frac{1}{r^{1-2\epsilon}}
\label{V0}
\end{equation}
(the power of $r$ is obvious from dimensions counting).
Here and below $\kappa_i = 1 + \mathcal{O}(\varepsilon)$
are some normalization factors (we don't need their exact form).

\begin{figure}[ht]
\begin{center}
\begin{picture}(49,20)
\put(24.5,11.5){\makebox(0,0){\includegraphics{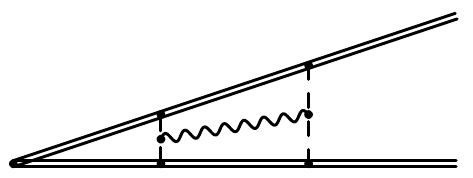}}}
\put(2,0){\makebox(0,0)[b]{0}}
\put(17,0){\makebox(0,0)[b]{$t_1$}}
\put(32,0){\makebox(0,0)[b]{$t_2$}}
\put(47,0){\makebox(0,0)[b]{$T$}}
\end{picture}
\end{center}
\caption{The first transverse-gluon contribution.}
\label{F:D}
\end{figure}
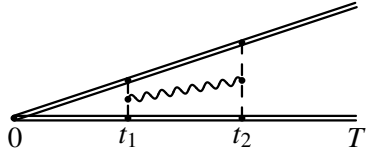

Transverse gluons interact only with Coulomb ones,
but not with static quarks.
The first transverse-gluon contribution is shown in Fig.~\ref{F:D}.
Here $T$ is an infrared cutoff.
We use the method of regions to analyze this contribution.
In the ultrasoft region $t_1 \sim t_2 \sim t_2-t_1$;
Coulomb gluons characteristic momentum is $q \sim 1/(ut_{1,2})$,
and the transverse gluon characteristic momentum is $k \sim 1/t_{1,2} \ll q$.
In the soft region $t_2-t_1 \sim ut_{1,2}$, and $k \sim 1/(t_2-t_1) \sim q$.
To determine the coefficient of the logarithm in the $1/\delta$ term in $\Gamma_{\text{cusp}}$,
it turns out to be sufficient to consider the ultrasoft region~\cite{GS}.
Neglecting $k$ in the 3-gluon vertex,
we obtain in momentum and coordinate spaces
\begin{equation}
\raisebox{-10mm}{\begin{picture}(13,22)
\put(6,11){\makebox(0,0){\includegraphics{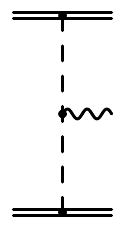}}}
\put(5,14){\makebox(0,0)[r]{$q$}}
\put(5,8){\makebox(0,0)[r]{$q$}}
\put(5,19){\makebox(0,0)[r]{$a_1$}}
\put(5,3){\makebox(0,0)[r]{$a_2$}}
\put(8,10){\makebox(0,0)[t]{0}}
\put(11,12){\makebox(0,0)[b]{$i$}}
\put(11,10){\makebox(0,0)[t]{$a$}}
\end{picture}}
= f^{a a_1 a_2} g_0^3 \frac{2 q^i}{(\vec{q}^2)^2}\,,\quad
\raisebox{-10mm}{\begin{picture}(13,22)
\put(6,11){\makebox(0,0){\includegraphics{us.pdf}}}
\put(5,19){\makebox(0,0)[r]{$\vec{r}$}}
\put(5,3){\makebox(0,0)[r]{0}}
\put(11,12){\makebox(0,0)[b]{$i$}}
\end{picture}}
= i f^{a a_1 a_2} \kappa_0 \frac{g_0^3}{4\pi} \frac{r^i}{r^{1-2\epsilon}}\,.
\label{us:vertex}
\end{equation}

The ratio of the Wilson line (Fig.~\ref{F:D}) to the one without the transverse-gluon correction is $1 + R_{\text{us}} + R_{\text{soft}}$.
The ultrasoft contribution is
\begin{equation}
R_{\text{us}} = \int_0^T dt_2 \int_0^{t_2} dt_1\,K(t_1,t_2)\,,
\label{WK}
\end{equation}
where
\begin{equation}
K(t_1,t_2) = \frac{1}{4} C_F C_A^2 \kappa_0^2 \frac{g_0^6}{(4\pi)^2}
\frac{r_1^i}{r_1^{1-2\epsilon}} \frac{r_2^j}{r_2^{1-2\epsilon}} D^{ij}(v(t_2-t_1))
\exp\left[- i \int_{t_1}^{t_2} dt\,\Delta V(ut)\right]
\label{K}
\end{equation}
($v=(1,\vec{0})$ is the 4-velocity of our small dipole).
During the time interval between $t_1$ and $t_2$,
the static quark--antiquark pair is in the adjoint color state instead of the singlet one,
and their leading-order interaction potential $V_o(r)$ is obtained from the expression
for the singlet potential $V(r)$~(\ref{V0}) by replacing the color factor $C_F$ with $C_F - C_A/2$.
Therefore, we get the integral of $\Delta V(r) = V_o(r) - V(r)$.
The characteristic sizes of the regions of the transverse gluon emission and absorption
are $\sim ut_{1,2}$;
we neglect them, so that this gluon propagates between the points $v t_1$ and $v t_2$:
\begin{equation}
D^{ij}(vt) = 8 (i/2)^{2\epsilon} \frac{\Gamma(2-\epsilon)}{3-2\epsilon}
\frac{t^{-2+2\epsilon}}{(4\pi)^{2-\epsilon}} \delta^{ij}\,.
\label{Dt}
\end{equation}
We obtain
\begin{equation}
K(t_1,t_2) = \frac{2}{3} C_F C_A^2 \kappa_1 \frac{g_0^6}{(4\pi)^4} u^{4\epsilon}
t_1^{2\epsilon} t_2^{2\epsilon} (t_2-t_1)^{-2+2\epsilon}
\exp\left[- \frac{i}{4} C_A \kappa_0 \frac{g_0^2}{4\pi}
\frac{t_2^{2\epsilon} - t_1^{2\epsilon}}{\epsilon u^{1-2\epsilon}}\right]\,.
\label{K2}
\end{equation}

Now we consider just a single Coulomb gluon exchange between $t_1$ and $t_2$:
\begin{equation}
K^{(1)}(t_1,t_2) = - \frac{i}{6} C_F C_A^3 \kappa_2 \frac{g_0^8}{(4\pi)^5}
\frac{t_1^{2\epsilon} t_2^{2\epsilon} (t_2^{2\epsilon}-t_1^{2\epsilon}) (t_2-t_1)^{-2+2\epsilon}}{\epsilon u^{1-6\epsilon}}\,.
\label{K1}
\end{equation}
Calculating the integral~(\ref{WK}) by the substitutions $t_1 = x t_2$ we obtain
\begin{equation}
\int_0^1 dx\,x^{2\epsilon} (1-x^{2\epsilon}) (1-x)^{-2+2\epsilon}
= \frac{\Gamma(1+2\epsilon)}{1-2\epsilon}
\left[3 \frac{\Gamma(1+4\epsilon)}{\Gamma(1+6\epsilon)}
- 2 \frac{\Gamma(1+2\epsilon)}{\Gamma(1+4\epsilon)}\right]
= 1 + \mathcal{O}(\varepsilon)\,,
\label{Intx}
\end{equation}
and
\begin{equation}
R_{\text{us}}^{(1)} = - \frac{i}{48} C_F C_A^3 \kappa_3 \frac{g_0^8}{(4\pi)^5} \frac{T^{8\epsilon}}{\epsilon^2 u^{1-6\epsilon}}\,.
%= 1 - \frac{i}{48} C_F C_A^3 \kappa_4 \frac{\alpha_s^4(\mu)}{4\pi} \frac{(\mu T)^{8\epsilon}}{\epsilon^2 u^{1-6\epsilon}}\,.
\label{Wres}
\end{equation}

The soft contribution is nearly local in time ($t_2-t_1 \sim u t_{1,2} \ll t_{1,2}$),
and can be described by a soft potential.
For a single coulomb exchange between $t_1$ and $t_2$, it is
\begin{equation}
V_{\text{soft}}^{(1)}(r) = c C_F C_A^3 \frac{g_0^8}{r^{1-8\varepsilon}}
\label{Vs1}
\end{equation}
by counting dimensions, so that
\begin{equation}
R_{\text{soft}}^{(1)} = - i \int_0^T dt\,V_{\text{soft}}^{(1)}(ut)
= - i c C_F C_A^3 \frac{g_0^8 T^{8\varepsilon}}{8 \varepsilon u^{1-8\varepsilon}}\,.
\label{Rs1}
\end{equation}
The double pole $1/\varepsilon^2$ should cancel
in $R^{(1)} = R_{\text{us}}^{(1)} + R_{\text{soft}}^{(1)}$;
this fixes the $1/\varepsilon$ term in $c$, and we obtain
\begin{equation}
R^{(1)} = - \frac{i}{48} C_F C_A^3 \frac{g_0^8 T^{8\varepsilon}}{(4\pi)^5}
\frac{\kappa_3 u^{6\varepsilon} - \kappa_4 u^{8\varepsilon}}{\varepsilon^2 u}
= \frac{i}{24} C_F C_A^3 \frac{\alpha_s^4(\mu) (\mu T)^{8\varepsilon}}{4\pi}
\frac{\log u + \text{const}}{\varepsilon u}\,.
\label{R1}
\end{equation}
This leads to the following contribution to $\Gamma_{\text{cusp}}$~\cite{GS}:
\begin{equation}
\Delta \Gamma_{\text{cusp}} = - \frac{i}{3} C_F C_A^3 \frac{\alpha_s^4}{4\pi}
\frac{\log u + \text{const}}{u}\,.
\label{Gamma:M}
\end{equation}
Finally, analytically continuing it to Euclidean space
($\varphi_E = \pi + i\varphi_M$, $\varphi_M = u$),
we obtain
\begin{equation}
\Delta \Gamma_{\text{cusp}}(\pi-\delta) = - \frac{1}{3} C_F C_A^3 \frac{\alpha_s^4}{4\pi}
\frac{\log\delta + \text{const}}{\delta}\,.
\label{Gamma:E}
\end{equation}

\textbf{Acknowledgements}. 
AG thanks N.~Brambilla and A.~Vairo for discussions of $\Gamma_{\text{cusp}}(\pi-\delta)$ (Sect.~\ref{delta}).
This work was supported in part by the Russian Ministry of science and higher education,
by the research training group GRK Symmetry Breaking (DFG/GRK 1581),
by the Deutsche Forschungsgemeinschaft through the project ``Infrared and threshold effects in QCD'',
by a GFK fellowship and by the PRISMA cluster of excellence at JGU Mainz.
The authors gratefully acknowledge the computing time granted on the supercomputer Mogon at JGU Mainz.

\end{document}